\documentclass[preprint]{aastex63}

\usepackage[version=4]{mhchem}

\usepackage{comment}
\includecomment{linenomath*}

\definecolor{darkorange}{RGB}{179,98,0}

\received{\today}
\revised{NaN}
\accepted{NaN}

\submitjournal{ApJL}

\shorttitle{Cyanoacetylene on GJ 1132 b}
\shortauthors{Rimmer et al.}

\graphicspath{{./}{figures/}}

\begin{document}

\title{Detectable Abundance of Cyanoacetylene (\ce{HC_3N}) Predicted on Reduced Nitrogen-Rich Super-Earth Atmospheres}

\correspondingauthor{Paul B. Rimmer}
\email{pbr27@cam.ac.uk}

\author[0000-0002-7180-081X]{Paul B. Rimmer}
\affiliation{Department of Earth Sciences, University of Cambridge, Downing St, Cambridge CB2 3EQ, United Kingdom}
\affiliation{Cavendish Laboratory, University of Cambridge, JJ Thomson Ave, Cambridge CB3 0HE, United Kingdom}
\affiliation{MRC Laboratory of Molecular Biology, Francis Crick Ave, Cambridge CB2 0QH, United Kingdom}

\author[0000-0001-7031-8039]{Liton Majumdar}
\affiliation{School of Earth and Planetary Sciences, National Institute of Science Education and Research, HBNI, Jatni 752050, Odisha, India}

\author[0000-0003-1143-0877]{Akshay Priyadarshi}
\affiliation{School of Physical Sciences, National Institute of Science Education and Research, HBNI, Jatni 752050, Odisha, India}

\author{Sam Wright}
\affiliation{Department of Physics and Astronomy, University College London, Gower Street, WC1E 6BT London, UK}

\author{S. N. Yurchenko}
\affiliation{Department of Physics and Astronomy, University College London, Gower Street, WC1E 6BT London, UK}


\begin{abstract}
We predict that cyanoacetylene (\ce{HC_3N}) is produced photochemically in the atmosphere of GJ 1132 b in abundances detectable by the James Webb Space Telescope (JWST), assuming that the atmosphere is hydrogen dominated and rich in molecular nitrogen (\ce{N2}), methane (\ce{CH4}) and hydrogen cyanide (\ce{HCN}), as described by \citet{Swain2021}. First, we construct line list and cross-sections for \ce{HC_3N}. Then we apply these cross-sections and the model atmosphere of \citet{Swain2021} to a radiative transfer model in order to simulate the transmission spectrum of GJ 1132 b as it would be seen by JWST, accounting for the uncertainty in the retrieved abundances. We predict that cyanoacetylene features at various wavelengths, with a clear lone feature at 4.5 $\mu$m, observable by JWST after one transit. This feature persists within the $1-\sigma$ uncertainty of the retrieved abundances of \ce{HCN} and \ce{CH_4}. The signal is detectable for stratospheric temperatures $\lesssim 600$ K and moderate stratospheric mixing ($10^6 \, {\rm cm^2 \, s^{-1}} \lesssim K_{zz} \lesssim 10^8 \, {\rm cm^2 \, s^{-1}}$). Our results also indicate that \ce{HC_3N} is an important source of opacity that future retrieval models should consider.
\end{abstract}


\section{Introduction} 
\label{sec:intro}

Cyanoacetylene (\ce{HC_3N}) is a linear molecule composed of two triple-bonded carbons bound at one end with a hydrogen atom and at the other with a nitrile group: carbon and nitrogen joined by a triple-bond. $\ce{H-C\!\equiv\! C-C\!\equiv\! N}$. Because of its two energetic bonds, it is both physically stable and highly reactive in aqueous media \citep{Ferris1968}. It acts as both a molecular backbone and a source of chemical energy for prebiotic chemical synthesis \citep{Powner2009,Becker2019}. It can be produced from radical reactions between the cyano radical and acetylene and is a major product of the Miller-Urey synthesis \citep{Miller1953,Sanchez1966}. There is extensive discussion among prebiotic chemists about whether \ce{HC_3N} is prebiotically plausible \citep{Orgel2002}.

\ce{HC_3N} has been observed in the atmosphere of Titan from the ground, both in the gas phase \citep{Bezard1992}, and condensed into crystalline aerosol \citep{Khanna2005}. According to retrieval from ALMA observations, the abundance of \ce{HC_3N} in the atmosphere of Titan is at about 100 ppb at 0.1 to 1 mbar, and drops by orders of magnitude with increasing pressure \citep{Thelen2019}. 

Experimentally, the IR intensities  of \ce{HC_3N} was studied in  \citet{06BeJoRa.HC3N,15DoGrLa.HC3N}. High-resolution spectroscopic analyses were reported by \citet{17BiFiJa.HC3N,21JiMeTa.HC3N}. 

\citet{Swain2021} analyzed of Hubble data of GJ 1132 b transmission spectra and find 0.5\% concentrations of both \ce{HCN} and \ce{CH_4} in an atmosphere with low mean molecular mass, implying the atmosphere is hydrogen-dominated. \citet{Swain2021} report that the atmosphere may be volcanic, and their photochemical models predict significant amounts of \ce{HC_3N} in the upper atmosphere. The analysis of \citet{Swain2021} has been challenged by \citet{Mugnai2021} and \citet{Libby2021}, who find no evidence of molecular features in the Hubble data. The analysis of \citet{Swain2021} and response by \citet{Mugnai2021} and \citet{Libby2021} are the latest in an elaborate and conflicting literature attempting to determine the presence and nature of an atmosphere of GJ 1132 b \citep[e.g.][]{Schaefer2016,Southworth2017,Diamond2018}.

Regardless of whether GJ 1132 b has a reduced atmosphere, there is theoretical support for the persistence of volcanically derived highly reduced atmospheres on rocky exoplanets, owing to the persistence of elemental iron in the mantle \citep{Lichtenberg2021}. It is possible that many Super Earth atmospheres resemble a warm Titan composition.

For this paper, we assume that the atmosphere reported by \citet{Swain2021} is the atmosphere of GJ 1132 b. Based on this assumption, we predict that \ce{HC_3N} is present in the upper atmosphere of GJ 1132 b in abundances detectable by the James Webb Space Telescope (JWST), and is an indication of a reduced atmosphere with a significant fraction of chemically active nitrogen and carbon species. In Section \ref{sec:methods}, we describe the method for constructing a line list of \ce{HC_3N} and the radiative transfer model and synthetic JWST pipeline used to calculate the transmission spectra. We present our results in Section \ref{sec:results}, and discuss the promise and challenges for observing \ce{HC_3N} on GJ 1132 b, and the application of this work beyond GJ 1132 b, in Section \ref{sec:discussion}.

\section{Methods}
\label{sec:methods}

In order to make predictions about \ce{HC_3N} in the atmosphere of GJ 1132 b, we first predict the atmospheric composition as a function of atmospheric height (Section \ref{sec:atmosphere}) and then develop a line list and cross-sections for \ce{HC_3N} (Section \ref{sec:linelist}), to model the transmission spectrum of GJ 1132 b (Section \ref{sec:spectra}). Finally, we simulate how that spectrum will look if observed by JWST (also Section \ref{sec:spectra}).

\subsection{Atmospheric Model}
\label{sec:atmosphere}

\citet{Swain2021} use the \textsc{Argo} model \citep{Rimmer2016} with the \textsc{Stand2020} chemical network \citep{Rimmer2019,Rimmer2021} and fixed surface conditions to predict atmospheric chemical profiles. \textsc{Argo} is a Lagrangian code that solves the photochemistry-transport equation:
\begin{equation}
\dfrac{dn_{\ce{X}}}{dt} = P_X - L_Xn_{\ce{X}} - \dfrac{\partial \Phi_X}{\partial z},
\end{equation}
where $n_{\ce{X}}$ (cm$^{-3}$) is the number density of species \ce{X}, $t$ (s) is time, $P_X$ (cm$^{-3}$ s$^{-1}$) is the production rate of \ce{X}, $L_X$ (s$^{-1}$) is the destruction rate of \ce{X}, and $\partial \Phi_X/\partial z$ (cm$^{-3}$ s$^{-1}$) accounts for the vertical transport. 

We take the chemical profiles of \citet{Swain2021} as given. This is a self-consistent photochemical model atmosphere, and so the predictions of each of the species are interconnected. Such an atmosphere cannot be rich in carbon dioxide, for example, because carbon dioxide is not a predicted thermochemical or photochemical product in such a reduced atmosphere. We also explore the sensitivity of \ce{HC_3N} to the concentrations of \ce{HCN} and \ce{CH_4}. To do this, we take the $\pm 1-\sigma$ errors for the retrieved abundances of \ce{HCN} and \ce{CH_4}, from \citet{Swain2021}, and apply or model to predict the effect on \ce{HC_3N}.

\subsection{Line list and cross sections / k-tables for Cyanoacetylene}
\label{sec:linelist}

There existed no comprehensive IR line list for \ce{HC_3N} applicable for a broad range of temperatures. The experimental and theoretical information is rather scarce for \ce{HC_3N}, especially in IR,  only for the room temperature  \citep{06BeJoRa.HC3N,07JoBeFa.HC3N,15DoGrLa.HC3N,17BiFiJa.HC3N,21JiMeTa.HC3N}. For this study, we have combined  spectroscopic data available for \ce{HC_3N} in the literature to construct a line list as a best estimate of the opacity of \ce{HC_3N} in IR. Our synthetic \ce{HC_3N}  line list  covers the wavelength  range  from 2.5 to 10~$\mu$m and contains the fundamental $\nu_1, \nu_2$, $\nu_3$ and overtones $2\nu_5$, $2\nu_6$, $3\nu_5$, $\nu_1+\nu_7$, $\nu_2+\nu_4$, $\nu_2+\nu_6$, $\nu_2+\nu_7$, $\nu_3+\nu_6$, $\nu_3+\nu_7$, $\nu_4+\nu_7$, $\nu_4+\nu_7$, $\nu_5+2\nu_7$, $\nu_5+\nu_6$ bands from this region. Only transitions from the ground vibrational state are included. 

We used the program PGOPHER \citep{PGOPHER} to generate energies and Einstein coefficients of \ce{HC_3N}. Spectroscopic constants of $2\nu_6$,  $\nu_5+\nu_7$, $\nu_6+\nu_7$, $\nu_6+2\nu_7$,  $\nu_1$, $\nu_2$, $\nu_3$, $\nu_1+\nu_7$, $\nu_2+\nu_7$ and $\nu_3+\nu_7$ were taken from the high-resolution IR studies by \citet{17BiFiJa.HC3N,21JiMeTa.HC3N}. Other states of \ce{HC_3N} from this IR region   have  not been characterized experimentally. Their spectroscopic constants were estimated using the constants of the same symmetry and were adjusted to visually  agree with the IR spectrum from \citet{06BeJoRa.HC3N}. The corresponding transition-dipole moments were extracted from the \textit{ab initio} work by  \citet{20DaPo.HC3N} computed at the high level of theory CCSD(T)-F12. 

\begin{figure}
\plotone{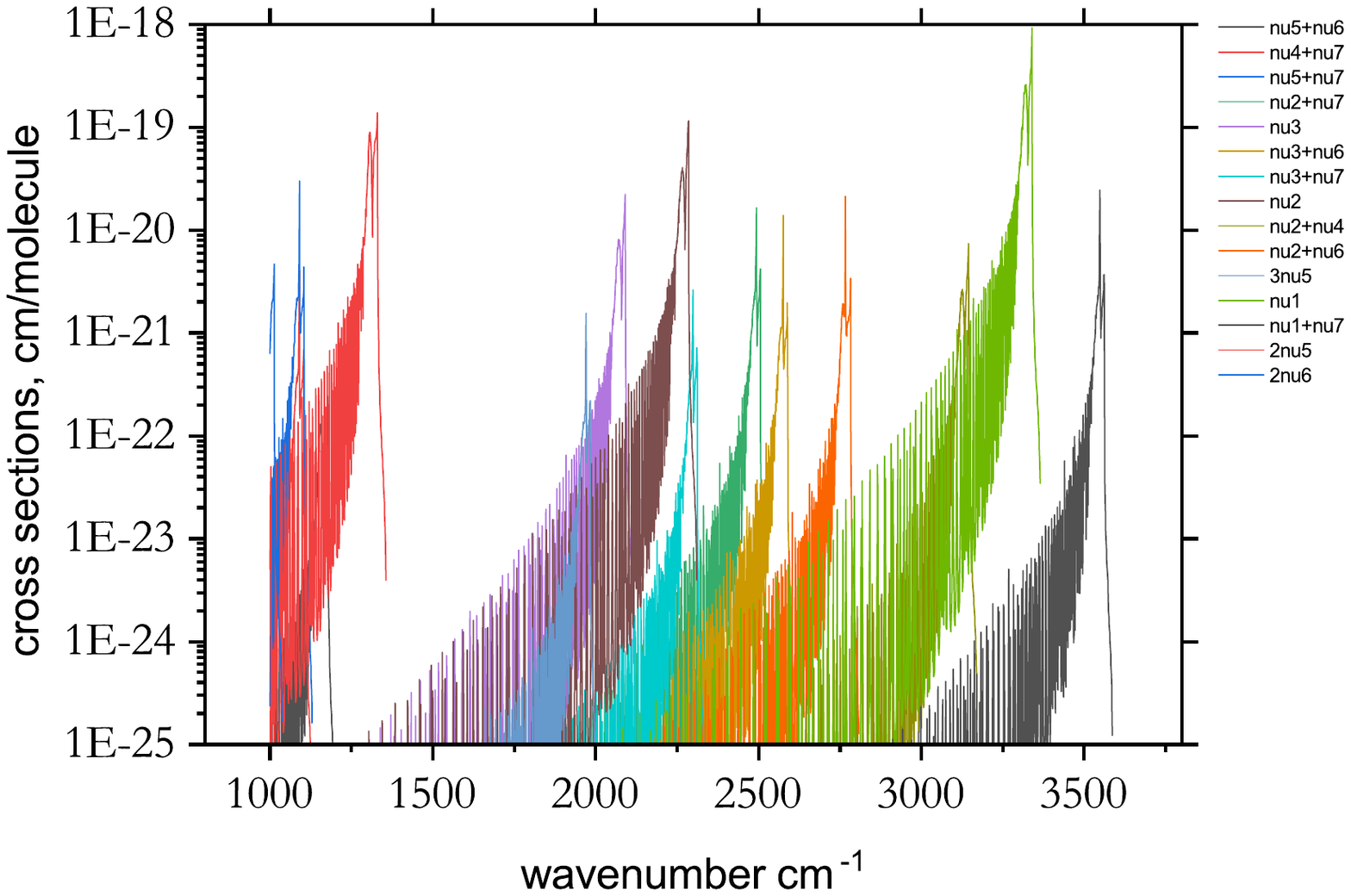}
\caption{Absorption cross-sections of HC$_3$N at $T=400$~K: vibrational bands (fundamentals and overtones)  used in the spectroscopic model are indicated in the legend. \label{fig:bands}}
\end{figure}

A set of temperature dependent cross sections and $k$ tables were generated using a combination of the program ExoCross \citep{ExoCross} and the Python library \verb|Exo_k| \citep{Leconte_2020} on the corresponding grids used in PetitRADTRANS \citep{molliere2019petitradtrans}. 
Figure~\ref{fig:bands} illustrates the vibrational bands used to generate the line list for HC$_3$N. 

\subsection{Predicted Transmission Spectra}
\label{sec:spectra}
From the atmospheric model, planetary physical parameters, and line-opacities of different molecules; synthetic transmission spectrum was modelled using  `PetitRADTRANS' \footnote{https://petitradtrans.readthedocs.io/en/latest/} \citep{molliere2019petitradtrans} both for a clear and cloudy atmosphere. The atmosphere was considered to be a composition of \ce{HC_3N}, \ce{HCN}, \ce{CH_4}, \ce{C_2H_2}, \ce{CO}, \ce{CO_2}, \ce{H_2O}, \ce{N_2O}, and \ce{NH_3}. Rayleigh scattering from \ce{H_2}, \ce{He}, and \ce{N_2} was considered, while Collision-induced absorption (CIA) opacities for \ce{H_2}-\ce{H_2}, and \ce{H_2}-\ce{He} were included. A correlated-k approximation mode was used, which resulted in a spectrum of Spectral Resolution $\lambda/\Delta\lambda=1000$. To model a cloudy atmosphere, a gray cloud deck was considered at 0.01 bar, which added an opaque cloud deck to the absorption opacity at the given pressure.

To study and analyse the detectability of molecular absorption features in the spectrum from JWST both for a clear and cloudy atmosphere, a noise simulator PandExo \citep{batalha2017pandexo} package was used. Following the details of modelling as mentioned in \citet{zilinskas2020atmospheric}, NIRCam grisms in F322W2 (2.4-4 $\mu m$) and F444W (4-5 $\mu m$) mode, and the MIRI LRS instrument in Slitless mode (5-12 $\mu m$) were used to imitate the spectrum. Constant noise levels of 30 ppm for NIRCam and 50 ppm for MIRI LRS were taken with a saturation limit of 80\% of full well capacity.

\section{Results} 
\label{sec:results}

Here we present our results for the \ce{HC_3N} chemistry in the atmosphere of GJ 1132 b, the model transmission spectrum for GJ 1132 b, and the synthetic JWST spectra.

\begin{figure}
\plottwo{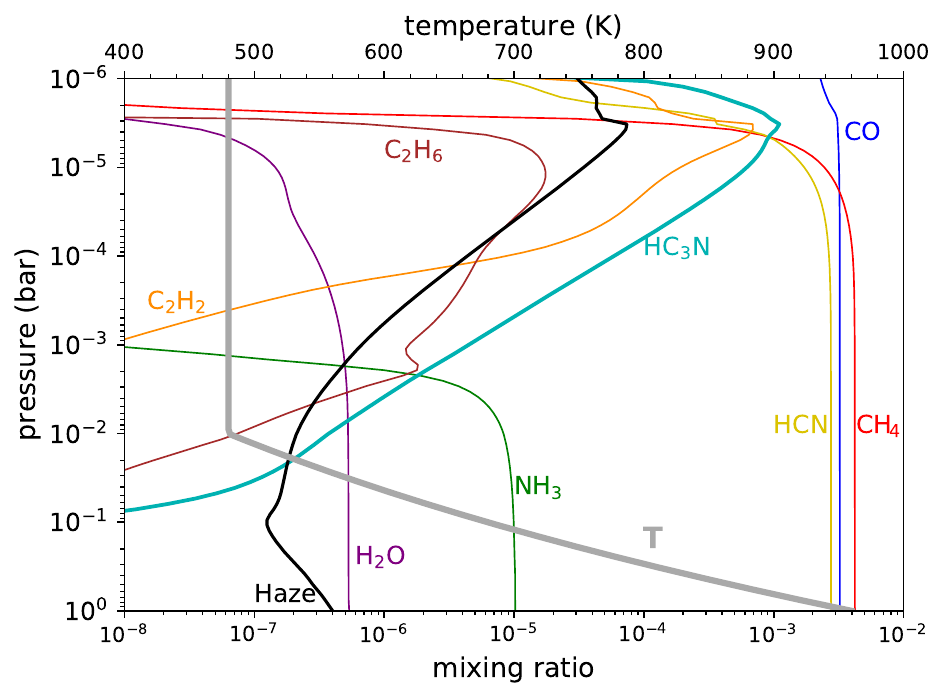}{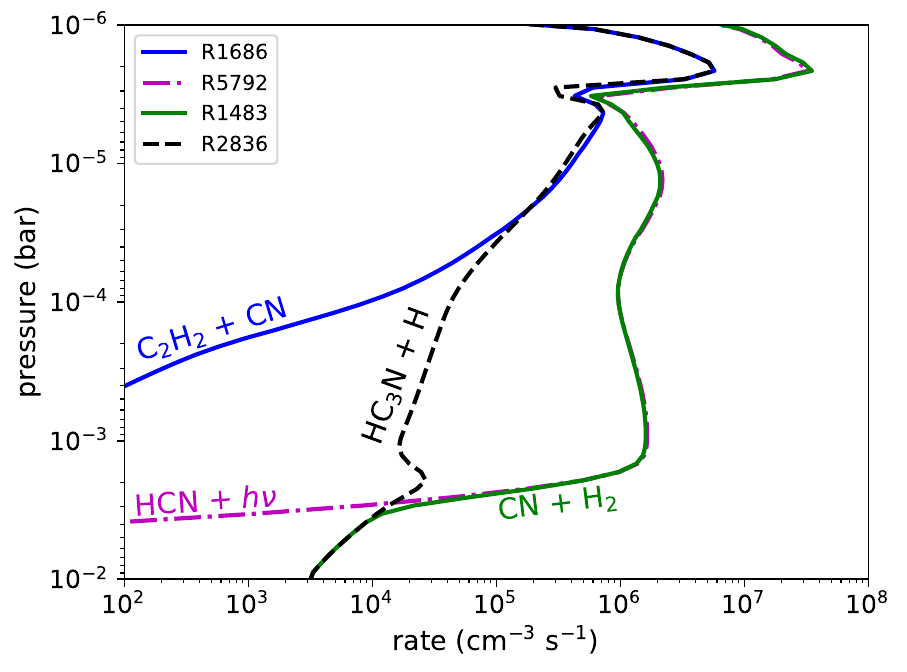}
\caption{{\bf Left:} Chemical profiles for GJ 1132 b, as mixing ratios as a function of atmospheric pressure (bar), highlighting \ce{HC_3N}. Taken from \citet[][their Fig. 11]{Swain2021}, with permission. {\bf Right:} Rates, in units of cm$^{-3}$ s$^{-1}$, for critical reactions for the formation and destruction of \ce{HC_3N} and its precursors, as a function of atmospheric pressure (bar). \label{fig:profiles}}
\end{figure}

The model atmosphere of GJ 1132 b sufficient to explain the observations from \citet{Swain2021} which involves a surface chemistry set by degassing primarily of \ce{H_2}, \ce{He}, \ce{N_2}, \ce{HCN}, \ce{CH_4} and \ce{CO}. Chemical profiles are shown in Figure \ref{fig:profiles}.

A major photochemical product of \ce{CH_4} is ethane (\ce{C_2H_6}), which is then dehydrogenated to acetylene (\ce{C_2H_2}), which at the high temperatures of the atmosphere of GJ 1132 b can react directly with the photodissociation product of \ce{HCN} to form \ce{HC_3N}. The total reaction proceeds as follows:
\begin{align}
\ce{HCN} + h\nu &\rightarrow \ce{CN} + \ce{H}, & \mathrm{R5792};\notag\\
2 \big(\ce{CH_4} + \ce{H} &\rightarrow \ce{CH_3} + \ce{H_2}\big), & \mathrm{R1520};\notag\\
\ce{CH_3} + \ce{CH_3} + \ce{M} &\rightarrow \ce{C_2H_6} + \ce{M}, & \mathrm{R143};\notag\\
\ce{C_2H_6} + \ce{H} &\rightarrow \ce{C_2H_5} + \ce{H_2}, & \mathrm{R1549};\notag\\
\ce{C_2H_5} + \ce{H} &\rightarrow \ce{C_2H_4} + \ce{H_2}, & \mathrm{R1538};\notag\\
\ce{C_2H_4} + h\nu &\rightarrow \ce{C_2H_2} + \ce{H} + \ce{H}, & \mathrm{R5841};\notag\\
\ce{CN} + \ce{C_2H_2} &\rightarrow \ce{HC_3N} + \ce{H}, & \mathrm{R1686};\notag\\
----&------\notag\\
\ce{HCN} + 2\ce{CH4} &\rightarrow \ce{HC_3N} + 4\ce{H2}.
\end{align}
Cyanoacetylene is destroyed by reaction with \ce{H} (R2386 the reverse of R1686), and this simply restores the cyano radical (\ce{CN}) and acetylene. This reaction has a barrier of $\approx 9500$ K, and this will be important for the steep temperature sensitivity of upper atmospheric cyanoacetylene, discussed below. The dominant reaction destroying \ce{CN} is R1483:
\begin{equation}
\ce{CN} + \ce{H_2} \rightarrow \ce{HCN} + \ce{H},
\end{equation}
which has a barrier of $2370\,{\rm K}$. The acetylene undergoes hydrogenation back to \ce{C_2H_5}, which then reacts with a hydrogen atom and breaks apart into two \ce{CH_3} radicals. At low temperatures, these would recombine to form ethane, but at high temperatures, the \ce{CH_3} will react with \ce{H_2} to reform \ce{CH4}. The significant reactions are shown as functions of atmospheric pressure in Figure \ref{fig:profiles}.

\begin{figure}[h!]
\plottwo{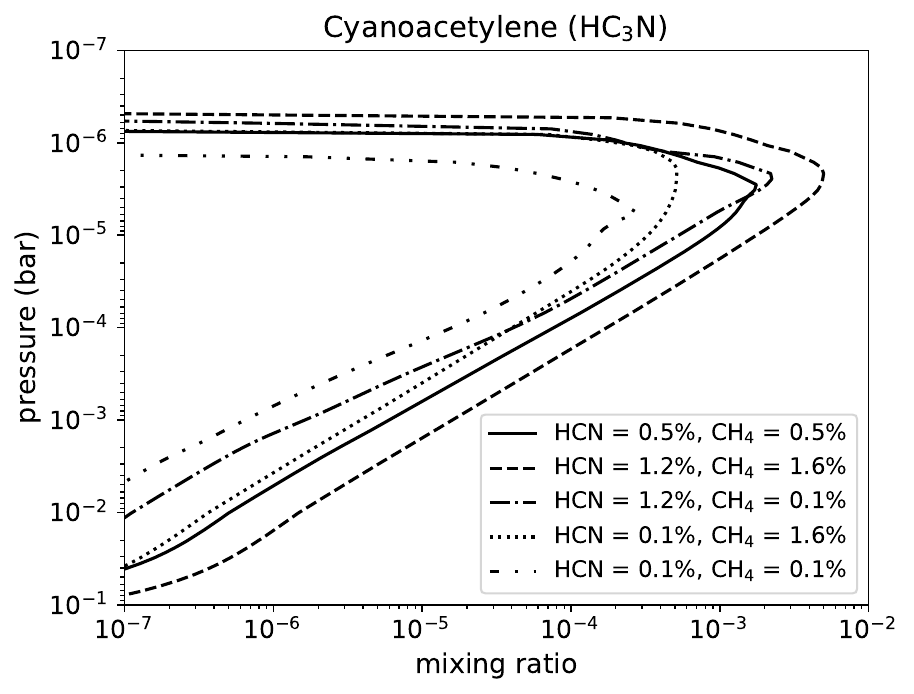}{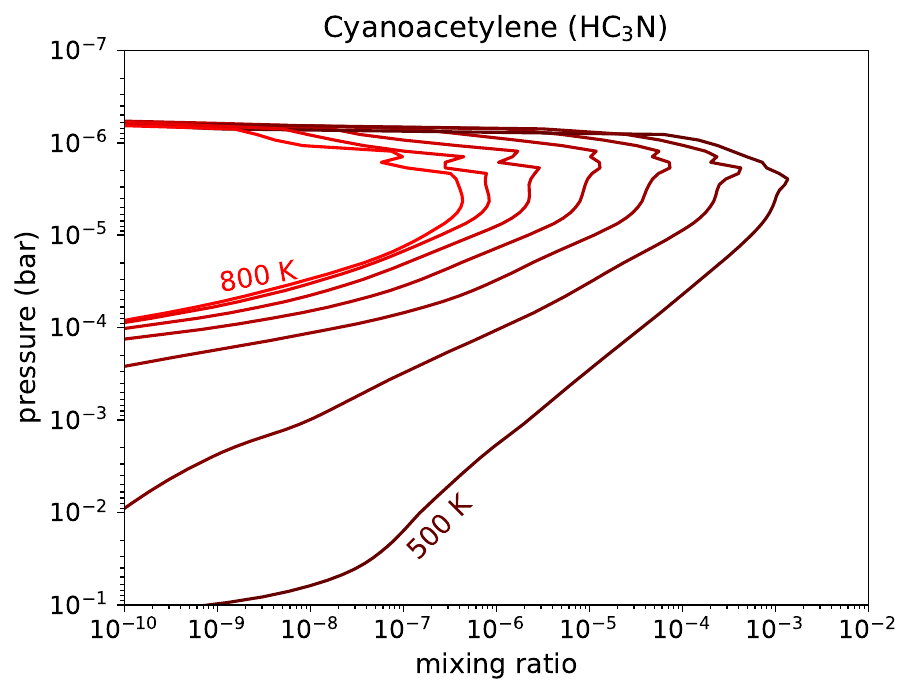}
\caption{Cyanoacetylene (\ce{HC_3N}) mixing ratio as a function of atmospheric pressure (bar). {\bf Left:} Predicted for a 10\% \ce{N_2}, 90\% \ce{H_2} atmosphere with the retrieved abundances and errors of \citet{Swain2021}: 0.5\% + 0.7\% - 0.4\% \ce{HCN} and 0.5\% + 1.1\% - 0.4\% \ce{CH_4}. \ce{C_2H_2} is set to 100 ppm. {\bf Right:} Predicted for 10\% \ce{N_2}, 90\% \ce{H_2} 0.5\% \ce{HCN} and \ce{CH_4}, as a function of stratospheric temperature. Colors proceed from dark red, 500 K, to light red, 800 K, in 50 K steps.
\label{fig:sensitivity}}
\end{figure}

We explore how \ce{HC_3N} varies as \ce{HCN} and \ce{CH_4} are varied. We vary both \ce{HCN} and \ce{CH_4} over the $1-\sigma$ retrieved abundances of \citet{Swain2021} and plot the predicted \ce{HC_3N} profiles in Figure \ref{fig:sensitivity}. \ce{HCN} abundance determines the peak of the \ce{HC_3N} profile, with some influence from \ce{CH_4}. \ce{CH_4} also influences the overall shape of the \ce{HC_3N} profile.

We also explore how \ce{HC_3N} varies with temperature and chemical mixing, expressed by varying the eddy diffusion coefficient ($K_{zz}$, cm$^2$ s$^{-1}$). We performed a sensitivity analysis with fixed \ce{HCN} and \ce{CH_4} at $0.5\%$, varying stratospheric temperature from 300 - 1000 K (with fixed $K_{zz} = 10^7$ cm$^2$ s$^{-1}$), and eddy diffusion coefficient from $10^5 - 10^{10}$ cm$^2$ s$^{-1}$ (with fixed stratospheric temperature of 480 K). 

The absorption lines corresponding to HC$_3$N are very narrow, and a reason for that could be its abundance only in the upper atmosphere, i.e., at low pressure, where pressure broadening is minimal. To resolve these narrow absorption lines, we'd need high spectral resolution. At a Spectral Resolution of 100 and 1 transit, the peaks of HC$_3$N at around 4.5 $\mu m$ and 9 $\mu m$ nevertheless seem distinguishable. If we increase the number of transits to 4, we see only a slight change in the values of error bars at higher wavelength regions, but it does not noticeably affect the error bars near HC$_3$N features. Even after considering cloudy atmosphere, an HC$_3$N feature is fairly detectable at around 9 $\mu$m when we have high HCN and CH$_4$. In all cases except when HCN and CH$_4$ both are low, the 4.5 $\mu$m \ce{HC_3N} line is readily detectable. The absorption features of other major molecules
viz. CH$_4$ and HCN are very prominent and have broad absorption lines in the generated spectrum, which aligns very well with their high abundance of those molecules, specifically in the higher pressure region as shown in the plot of pressure vs. mixing ratio of species. At least one of the \ce{HC_3N} features is detectable for stratospheric temperature $\lesssim 600$ K, as shown in Figure \ref{fig:sensitivity}, and eddy diffusion coefficient between $10^6$ cm$^2$ s$^{-1}$ $\lesssim K_{zz} \lesssim$ $10^9$ cm$^2$ s$^{-1}$. If we decrease the spectral resolution to 50, we can still distinguish the \ce{HC_3N} feature at 4.5 $\mu$m for most of the models (except the model with $0.1\%$ HCN and $0.1\%$ \ce{CH_4}), whereas \ce{HC_3N} feature at 9 $\mu$m remains undetectable for most of the models (except the model with $1.2\%$ HCN and $1.6\%$ \ce{CH_4}). Thus, the spectral resolution of 100 is required to detect both the \ce{HC_3N} features simultaneously.

We see in Figure \ref{fig:sensitivity} that the mixing ratio of \ce{HC_3N} is strongly dependent on stratospheric temperature, and drops below 1 ppm in the upper atmosphere when $T \gtrsim 700$ K. The relative strength of these lines and others in the upper atmosphere is also affected by the temperature, and at high temperatures, $\gtrsim 600$ K, \ce{HC_3N} will be very difficult to detect. Above $\sim 700$ K, no observable features remain.

This temperature sensitivity can be explained by R2386 (discussed above):
\begin{equation}
\ce{HC_3N} + \ce{H} \rightarrow \ce{C_2H_2} + \ce{CN},
\end{equation}
The rate constant for this reaction between $450 - 800$ K is effectively $2.5 \times 10^{-10} \; {\rm cm^3 \, s^{-1}} \exp(-9500 \, {\rm K}/T)$, and production does not change by more than a factor of a few over this range, and so balancing production and destruction at $\sim 10^{-5}$ bar, one finds an analytic estimate of the peak mixing ratio of cyanoacetylene ($x(\ce{HC_3N})$):
\begin{equation}
x(\ce{HC_3N}) \approx 4 \times 10^{-12} \, e^{9500/T}, \; \; \; 450 \, {\rm K} \leq T \leq 800 \, {\rm K}.
\label{eqn:hc3n-T}
\end{equation}
This explains the chemical mechanism for the loss of cyanoacetylene at higher stratospheric temperatures.

\begin{figure}[h!]
\centering
\includegraphics[width=0.98\textwidth]{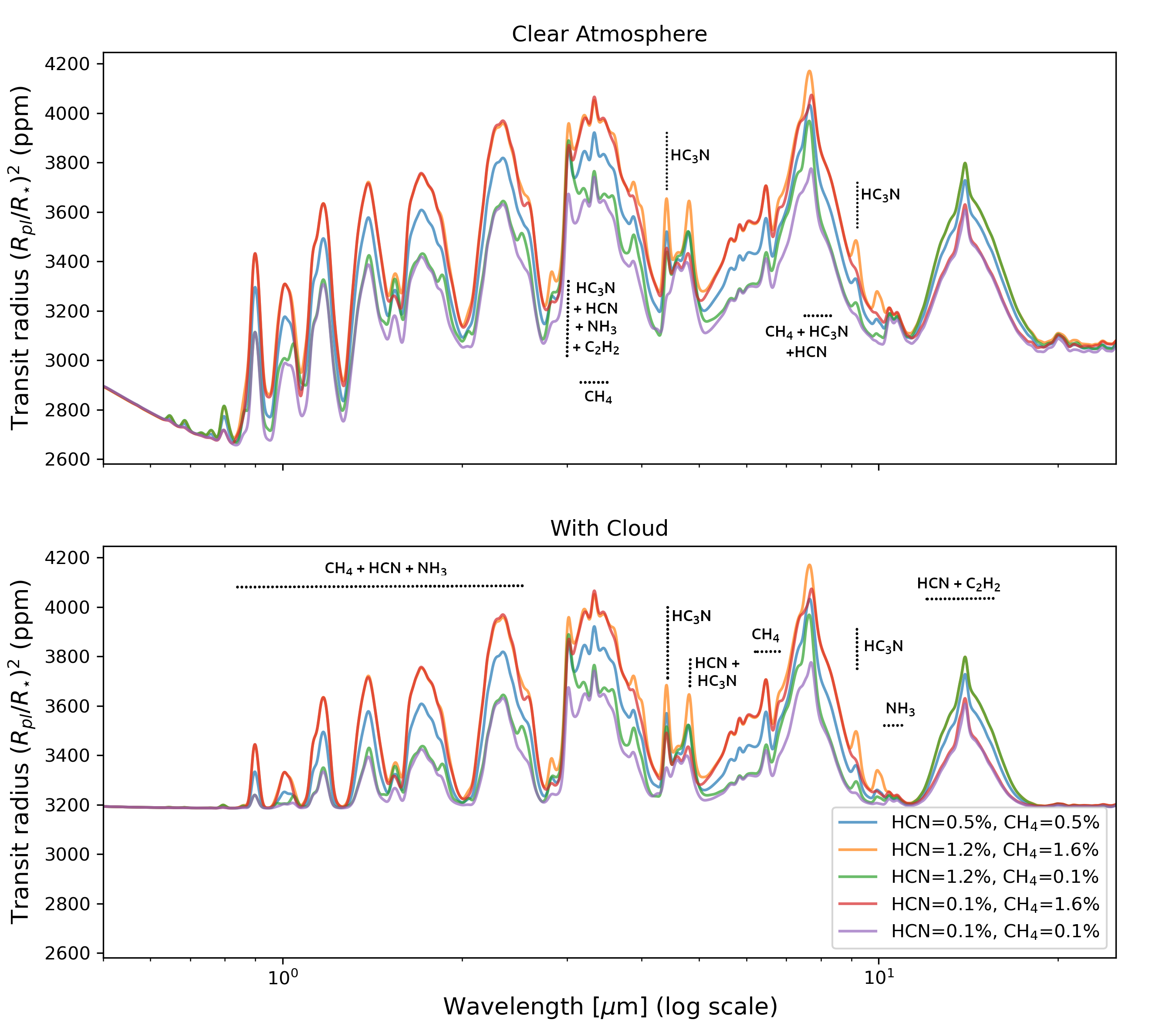}
\caption{Simulated transmission spectrum of GJ1132b using \textsc{Argo} photochemistry model and \textsc{PetitRADTRANS} radiative transfer package for different \ce{HCN} and \ce{CH_4} compositions. The cloud was modelled according to the power law function, considering a Gray cloud at P=0.01 bar. The spectrum is shown at a resolution of 100, which was generated by convolving the high resolution data (R=1000) with a Gaussian kernel of standard deviation=8.4506.}
\label{fig:transm}
\end{figure}

\begin{figure}[h!]
\centering
\includegraphics[width=0.9\textwidth]{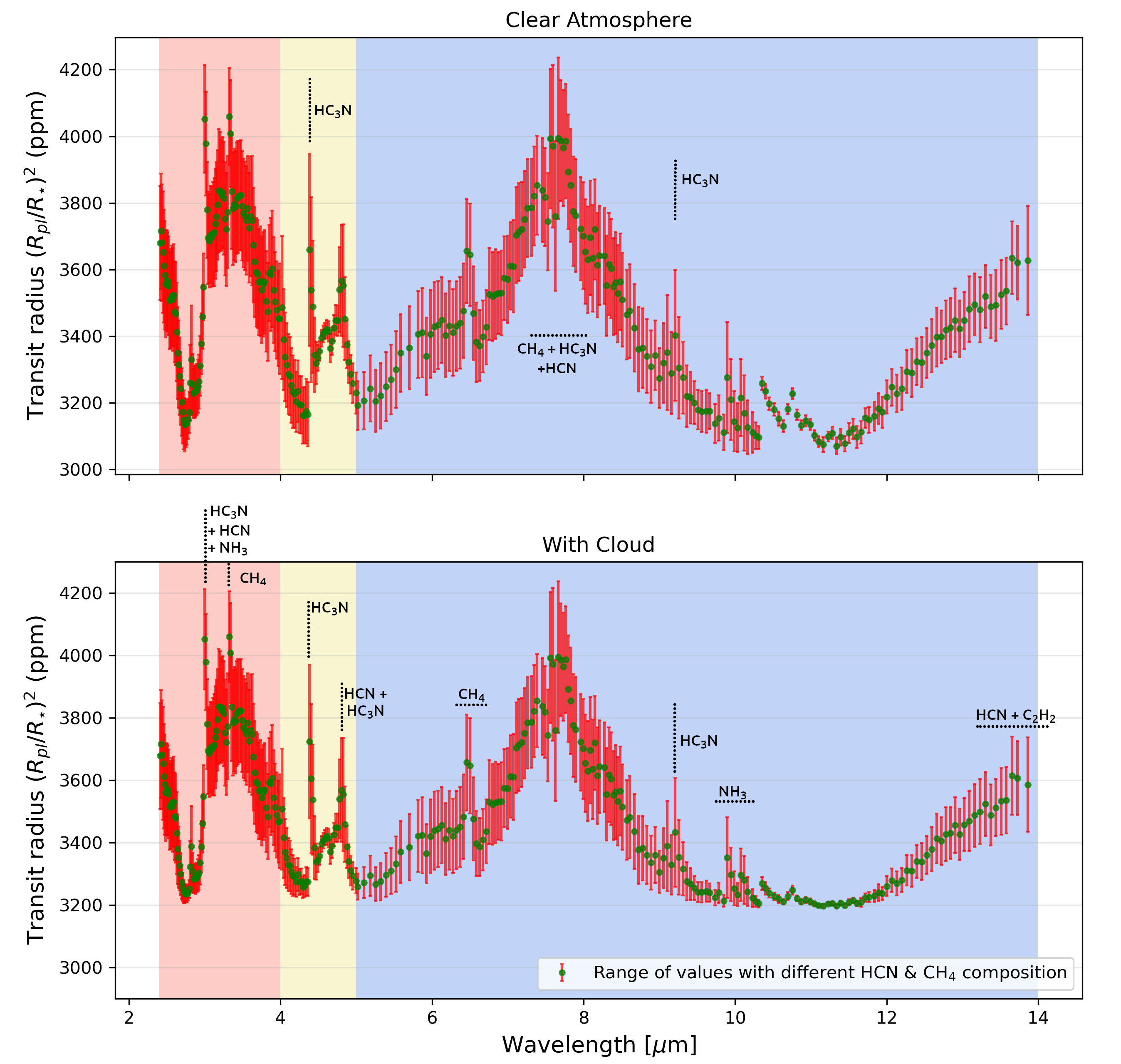}
\caption{The range of values in the JWST simulated observation using PandExo for different \ce{HCN} and \ce{CH_4} compositions, after one transit and with a resolution of 100. The cloud was modelled according to the power law function, considering a Gray cloud at P=0.01 bar. The three colors denote the three JWST modes and instruments used, which were NIRCam grism F322W2, NIRCam grism F444W, and MIRI LRS.}
\label{fig:jwst}
\end{figure}

\section{Discussion and Conclusion} 
\label{sec:discussion}

In this letter, we show that, if GJ 1132 b has observable amounts of \ce{CH_4} and \ce{HCN} in its atmosphere, as reported by \citet{Swain2021}, we predict it will have up to 0.5\% of \ce{HC_3N} in its upper atmosphere due to photochemistry. Applying radiative transfer and accounting for the properties of JWST, we predict that cyanoacetylene (\ce{HC_3N}) will be observable by JWST, given what we presently know about the molecule.

There is some uncertainty about the \ce{HC_3N} line list (limited coverage of vibrational bands), but these uncertainties are small compared to the uncertain composition of the atmosphere, with retrieved \ce{HCN} and \ce{CH_4} abundances spanning more than an order of magnitude and the predicted \ce{HC_3N} abundances spanning a factor of 30. These uncertainties were incorporated into the spectra and we found that spectral features of the molecules remain detectable at $\gtrsim 0.1$\% mixing ratios of \ce{HCN} and \ce{CH_4}, the low end of the $1-\sigma$ error bars for these species retrieved abundances. More work will be needed, both experimental and theoretical, in order to make any strong claim about the shape and strength of \ce{HC_3N} features for temperatures $\gg 500$ K, where otherwise energetically prohibitive and therefore yet unknown bands will be accessed, and the spectrum of \ce{HC_3N} could be significantly different. As we have found, however, \ce{HC_3N} abundance decreases to below detectable levels (concentrations of 10-100 ppm) at temperatures $\gtrsim 600$ K, at least in the reducing rocky planet atmosphere we explore. For our bulk atmospheric composition, \ce{HC_3N} may not be detectable at such high stratospheric temperatures because it will not survive in those temperatures, see Eq. (\ref{eqn:hc3n-T}). We found that \ce{HC_3N} concentrations were far less sensitive to the eddy diffusion coefficient, remaining detectable between values of $10^6$ cm$^2$ s$^{-1}$ and $10^{10}$ cm$^2$ s$^{-1}$.

Even if GJ 1132 b does not have \ce{HCN}, there are many planets, including, plausibly, many rocky planets that may host hydrogen-rich atmospheres rich in \ce{HCN} \citep{Tsiaras2016}. Super Earths also may host reduced atmospheres because of quenched mantle differentiation \citep{Lichtenberg2021}. For reduced atmospheres of rocky planets, \ce{HC_3N} is an important opacity source that retrieval models should take into account. 

By itself, \ce{HC_3N} cannot be uniquely identified by observing one or two spectral features. Rather, it will be presence in the context of a hydrogen-rich exoplanet atmosphere. The probability that \ce{HC_3N} is an explanation for a given spectral feature must be assessed in the context of other molecular constituents as indicated by the full transmission spectra, such as \ce{H_2} (inferred from the scale height), \ce{CH_4}, and \ce{HCN}. In this context, \ce{HC_3N} would be a further indicator of the presence of \ce{HCN}, and a probe into prebiotically relevant mechanisms taking place on uninhabitable planets. This may give a window into the time when our planet was hot and hydrogen-rich \citep{Genda2017}, forming molecules that later may have been essential to the origins of life.

\acknowledgments

The authors thank Mark Swain for helpful comments on the paper. P.~B.~R. thanks the Simons Foundation for funding (SCOL awards 599634). SY's work was supported by the STFC Project  ST/R000476/1 and 
by the European Research Council (ERC) under the European Union’s Horizon 2020 research and innovation programme through Advance Grant number 883830. S.W. was supported through the STFC UCL CDT in Data Intensive Science (grant No. ST/P006736/1). 

\vspace{5mm}

\software{\textsc{Argo} \citep{Rimmer2016}, \textsc{petitRADTRANS} \citep{molliere2019petitradtrans}, \textsc{PandExo} \citep{batalha2017pandexo}},
\verb|Exo_k|
\citep{Leconte_2020},
\textsc{ExoCross}
\citep{ExoCross}

\end{document}